\documentclass{webofc}
\hypersetup{colorlinks=true,urlcolor=teal,
                      linkcolor=teal,citecolor=teal}
\usepackage[varg]{txfonts}   
\usepackage[utf8]{inputenc} 
\usepackage{xurl}
\usepackage{amsfonts}       
\usepackage{nicefrac}       
\usepackage{microtype}      
\usepackage{amsmath}        
\usepackage{braket}            
\usepackage[nolist]{acronym} 
\usepackage{graphicx}
\usepackage[inline]{enumitem}
\usepackage{tikz}
\usepackage{layouts}
\usepackage{subcaption}
\usepackage{yquant}
\usepackage{csquotes}

\newcommand{\eg}{\emph{e.g.}\xspace}
\newcommand{\ie}{\emph{i.e.}\xspace}
\newcommand{\etal}{\emph{et al.}\xspace}

\newcommand{\FOURIER}{\textsc{Fourier}\xspace}

\DeclareMathOperator*{\argmin}{arg\,min}

\begin{document}

\begin{acronym}[TROLL]
  \acro{qc}[QC]{Quantum Computing}
  \acro{nisq}[NISQ]{Noisy Intermediate-Scale Quantum}
  \acro{nhep}[NHEP]{Nuclear and High-Energy Physics}
  \acro{qa}[QA]{Quantum Annealing}
  \acro{qaoa}[QAOA]{Quantum Approximate Optimisation Algorithm}
  \acro{qubo}[QUBO]{Quadratic Unconstrained Binary Optimisation}
  \acro{lhc}[LHC]{Large Hadron Collider}
  \acro{tr}[TR]{Track Reconstruction}
  \acro{sg}[SG]{Spectral Gap}
  \acro{ri}[RI]{Random Parameter Initialisation}
\end{acronym}

\def\chapterautorefname{Chap.}%
\def\sectionautorefname{Sec.}%
\def\subsectionautorefname{Sec.}%
\def\subsubsectionautorefname{Sec.}%
\def\paragraphautorefname{Par.}%
\def\tableautorefname{Tab.}%
\def\equationautorefname{Eq.}%
\def\figureautorefname{Fig.}%


\title{From Hope to Heuristic: Realistic Runtime Estimates for Quantum Optimisation in NHEP}

\author{\firstname{Maja} \lastname{Franz}\inst{1}\fnsep\thanks{\email{maja.franz@othr.de}} \and
  \firstname{Manuel} \lastname{Schönberger}\inst{1} \and
  \firstname{Melvin} \lastname{Strobl}\inst{2}\fnsep\thanks{\email{melvin.strobl@kit.edu}} \and
  \firstname{Eileen} \lastname{Kühn}\inst{2} \and
  \firstname{Achim} \lastname{Streit}\inst{2} \and
  \firstname{Pía} \lastname{Zurita}\inst{3} \and
  \firstname{Markus} \lastname{Diefenthaler}\inst{4} \and
  \firstname{Wolfgang} \lastname{Mauerer}\inst{1, 5}
}

\institute{%
  Technical University of Applied Sciences Regensburg, Germany
  \and Karlsruhe Institute of Technology, Germany
  \and Complutense University of Madrid, Spain
  \and Jefferson Lab, VA, USA
  \and Siemens AG, Technology, Munich, Germany%
}

\abstract{%
  \ac{nisq} computers, despite their limitations, present opportunities for near-term quantum advantages in \ac{nhep} when paired with specially designed quantum algorithms and processing units.
  This study focuses on core algorithms that solve optimisation problems through the quadratic Ising or quadratic unconstrained binary optimisation model, specifically quantum annealing and the \ac{qaoa}.

  In particular, we estimate runtimes and scalability for the task of particle \ac{tr}, a key computing challenge in \ac{nhep}, and investigate how the classical parameter space in \ac{qaoa}, along with techniques like a \FOURIER-analysis based heuristic, can facilitate future quantum advantages.
  The findings indicate that lower frequency components in the parameter space are crucial for effective annealing schedules, suggesting that heuristics can improve resource efficiency while achieving near-optimal results.
  Overall, the study highlights the potential of \ac{nisq} computers in \ac{nhep} and the significance of co-design approaches and heuristic techniques in overcoming challenges in quantum algorithms.
}

\maketitle

\acresetall

\section{Introduction}
\ac{nisq} computers, while limited by imperfections and small scale, hold promise for near-term quantum advantages in \ac{nhep} when coupled with co-designed quantum algorithms and special-purpose quantum processing units~\cite{franz_co-design_2024}.
Developing co-design approaches is essential for near-term usability, but inherent challenges exist due to the fundamental properties of \ac{nisq} algorithms~\cite{preskill_quantum_2018}.
In this contribution we therefore investigate the core algorithms, which can solve optimisation problems via the abstraction layer of a quadratic Ising model or general \ac{qubo}, namely \ac{qa} and the \ac{qaoa}.
As an example for a variety of applications in \ac{nhep} utilising \ac{qubo} formulations~\cite{pakhomchik_solving_2022}, we focus on \ac{tr}~\cite{okawa_quantum-annealing-inspired_2024,zlokapa_charged_2021}.
\ac{tr} is a good example of pattern recognition and optimisation in \ac{nhep}, and with TrackML~\cite{Kiehn_2019}, there exist curated datasets for algorithmic development along well-documented performance metrics.
While \ac{qa} and \ac{qaoa} do not inherently imply quantum advantage, \ac{qa} runtime for specific problems can be determined based on the physical properties of the underlying Hamiltonian, although it is a computationally hard problem itself~\cite{albash_adiabatic_2018}.
Our primary focus is on two key areas:
Firstly, we estimate runtimes and scalability for the common \ac{nhep} problem of \ac{tr} addressed via a \ac{qubo} formulation~\cite{bapst_pattern_2019}.
This analysis is conducted by identifying minimum energy solutions of intermediate Hamiltonian operators encountered during the annealing process.
Secondly, we investigate how the classical parameter space in the \ac{qaoa}, together with approximation techniques such as a \FOURIER-analysis based heuristic, proposed by Zhou~\etal~\cite{zhou_quantum_2020}, can help to achieve (future) quantum advantage, considering a trade-off between computational complexity and solution quality.

The remainder of this article is structured as follows:
In \autoref{sec:qa} we provide a brief introduction to \ac{qa} and explain how annealing schedules and energy levels link to \acp{sg}.
\autoref{sec:qaoa} then introduces \ac{qaoa}, how it can be used to derive annealing schedules, and what heuristics exist for doing so.
In \autoref{sec:trackrec}, we briefly introduce the task of \ac{tr} and describe our experimental setup with the results in \autoref{sec:exp}.
Finally, we conclude in \autoref{sec:concl}.
We ensure reproducibility~\cite{mauerer:22:q-saner}, by providing the code as well as the numerical results that we use in the figures of this work in Ref.~\cite{franz_2025_14921650}.
Notably, our experiments are performed using classical simulations of \ac{qaoa} and \ac{qa} with Qiskit~\cite{qiskit2024}.

\section{Quantum Annealing}
\label{sec:qa}
\ac{qa} is, roughly speaking, a restricted version of adiabatic quantum computing~\cite{albash_adiabatic_2018} which is used to find minimum energy solutions to a problem (cost) Hamiltonian \(\hat{H}_C\).
The solutions are obtained by an adiabatic transition from a system, prepared in the ground state of an initial Hamiltonian \(\hat{H}_{0}\), to \(\hat{H}_C\).
\ac{qa} is restricted to a certain subclass of all possible Hamiltonians~\cite{albash_adiabatic_2018}, which are equivalent to \ac{qubo} problems, to which any \textbf{NP} problem can be reduced~\cite{lucas_ising_2014}.
The overall time-dependent Hamiltonian of the annealing process is determined by the \enquote{protocol} \(f(t)\) that guides the transition between the Hamiltonians via
\begin{equation}
  \hat{H}_{QA}(t) = -\left(f(t)\hat{H}_C + (1-f(t))\hat{H}_{0}\right)\label{eq:QA}
\end{equation}

By default, D-Wave annealers~\cite{dwave} implement a linear transition given by
\(f(t)=s \coloneqq t/T\), where \(T\) is the chosen \emph{annealing time},
and \(t\in [0,T]\).
It is known that the minimum time \(T_{\text{min}}\) required for an adiabatic
transition depends on the spectrum of the Hamiltonian \(\hat{H}_{QA}\), in particular
the minimum energy gap (\ac{sg} \(\Delta_\text{min}\)) between ground state and first
excited state encountered when running the protocol given in \autoref{eq:QA} by \(T_\text{min} = \mathcal{O}(1/\Delta_\text{min}^{2})\).
Harder problems with longer required runtimes therefore comprise smaller minimum \acp{sg}.

\begin{figure}[ht]
  \begin{subfigure}{0.5\textwidth}
    \includegraphics[width=0.95\textwidth]{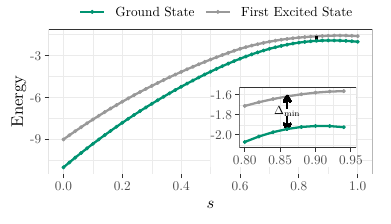}
  \end{subfigure}
  \begin{subfigure}{0.5\textwidth}
    \includegraphics[width=0.95\textwidth]{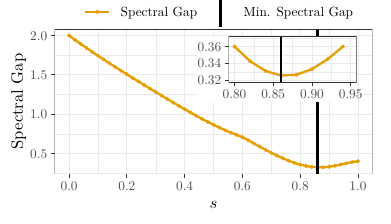}
  \end{subfigure}
  \caption{
    Left: Energy values for ground and first excited state of the Hamiltonian given in \autoref{eq:QA} over interpolation parameter $s$ for an example \ac{tr} problem instance considered in this work.
    Right: Corresponding \ac{sg} (\ie, the difference between energy levels of the first excited and the ground state).
    Lines do not provide an interpolation, but are only used to guide the eye; insets magnify the region around the minimal \ac{sg}.
  }
  \label{fig:energy_selected}
\end{figure}

Consider, as shown on the left-hand side of \autoref{fig:energy_selected}, the \(s\)-dependent discrete energy levels of the ground state and the first excited state of one of the \ac{tr} problem instances considered in this work (see \autoref{sec:trackrec}).
Physical intuition suggests that there are regions with a large gap between the two states, where the annealing process can proceed more rapidly without the encountering level transitions.
Conversely, in regions characterised by a small \ac{sg}, the annealing process must conducted at a slower rate to prevent the system from entering higher energy states, as this would result in sub-optimal or invalid solutions.
The minimum \ac{sg} is observed at around \(s=0.85\) and persists at low levels until \(s=1\).

The annealing schedule, denoted by $f(t)$, can be adapted to accommodate the properties of the problem under consideration. Intuitively, this means that during the first three-quarters of the time, the transition from Hamiltonian $\hat{H}_{0}$ to $\hat{H}_{C}$ can proceed swiftly, and that it needs to slow down in the last quarter.
However, givent the computational complexity of the energy spectrum of the Hamiltonian~(\autoref{eq:QA}) and its unavailability prior to the problem's formulation, alternative means of determining good schedules are necessary.

\section{Quantum Approximate Optimisation Algorithm}
\label{sec:qaoa}

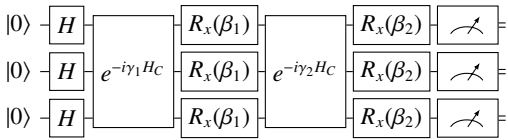
\begin{figure}[ht]
  \begin{minipage}[c]{0.6\textwidth}
    \begin{tikzpicture}[font=\small]
  \begin{yquant}
    qubit {$\ket{0}$} q[3];

    h q[0-2];
    box {$e^{-i\gamma_1 H_C}$} (q[0-2]);
    box {$R_x(\beta_1)$} q[0-2];
    box {$e^{-i\gamma_2 H_C}$} (q[0-2]);
    box {$R_x(\beta_2)$} q[0-2];
    measure q[0-2];

  \end{yquant}
\end{tikzpicture}
  \end{minipage}\hfill
  \begin{minipage}[c]{0.35\textwidth}
    \caption{\ac{qaoa} circuit for cost Hamiltonian $H_C$ and parameter vectors $\vec\beta$, $\vec\gamma$ of size $p=2$.}
    \label{fig:qaoa}
  \end{minipage}
\end{figure}

One such possibility is to use insights gained from \ac{qaoa} computations.
As \ac{qa}, the \ac{qaoa} is designed to optimise \ac{qubo} problems.
It employs a quantum circuit with $p \in \mathbb{N}$ layers of unitary operators defined by $2p$ parameters $\vec{\beta}, \vec{\gamma} \in \mathbb{R}^{p}$.
A \ac{qaoa} layer $j \in [1, p]$ comprises two unitaries:
\begin{equation}
  U_M(\beta_j) = e^{-i\beta_j\hat{H}_M} \quad\mathrm{and}\quad U_C(\gamma_j) = e^{-i\gamma_j\hat{H}_C}
\end{equation}
with $U_M$ representing mixer Hamiltonian $\hat{H}_M$, and $U_C$ based on the cost Hamiltonian $\hat{H}_C$, of which the ground state encodes the optimal solution to a given \ac{qubo} problem.
The mixer unitary $U_M$ typically consists of $\hat{X}$-rotations of size $\beta_j$ on each qubit, while the cost unitary $U_C$ uses single, or multi-qubit $\hat{Z}$-rotations of size $\gamma_j$.
The initial state $\ket{s}$ of the \ac{qaoa} algorithm is usually chosen as the ground state of $H_M$, in which each qubit is in an equal superposition of $\ket{0}$ and $\ket{1}$, prepared using a layer of Hadamard gates.
The repeated application of these layers results in the parameterised quantum state
\begin{equation}\label{eq:qaoa_parameterized_state}
  \ket{\gamma,\beta} = U_{M}(\beta_{p})U_{C}(\gamma_{p})\cdots U_{M}(\beta_{1})U_{C}(\gamma_{1})\ket{s},
\end{equation}
which corresponds to the discretised time evolution governed by the Hamiltonians $H_M$ and $H_C$.
A general example of a three-qubit \ac{qaoa} circuit with $p=2$ is illustrated in \autoref{fig:qaoa}.
It has been established that the quality of the approximation increases for a larger number of layers~\cite{farhi_quantum_2014}.
However, it is important to note that the overall solution quality is significantly influenced by the parameter values $\vec{\beta}$ and $\vec{\gamma}$.

\subsection{Annealing Schedule Derivation}
\label{sec:anneal_schedules}
It is known~\cite{zhou_quantum_2020} that the optimal parameter vectors $(\vec{\gamma}^{*}, \vec{\beta}^{*}) = \argmin_{\vec{\gamma}, \vec{\beta}} \bra{\vec{\gamma}, \vec{\beta}}H_C\ket{\vec{\gamma}, \vec{\beta}}$, where $\bra{\vec{\gamma}, \vec{\beta}}H_C\ket{\vec{\gamma}, \vec{\beta}}$ is the observed mean energy of the final quantum state, can be interpreted as (smooth) annealing path with total annealing time given by \(T=\sum_{i=1}^{p}(|\gamma_{i}^*|+|\beta_{i}^*|)\).
The path itself is constructed from supporting points \(f(t_{i}) = \frac{\gamma_{i}^{*}}{|\gamma_{i}^{*}|+|\beta_{i}^{*}|}\) that are connected via linear interpolation, with time \(t_{i}\) chosen as the mid-point of interval \(\gamma_{i}^{*}, \beta_{i}^{*}\).
However, finding \emph{optimal} parameters $\vec{\gamma^{*}}, \vec{\beta^{*}}$ is shown to be \textbf{NP}-hard~\cite{bittel_training_2021}, albeit for smaller depths, empirical and theoretical results ascertain that optimal values can be well approximated for many subject problems~\cite{bharti_noisy_2022}.

\subsection{Structure of the \ac{qaoa} parameter space}
\label{sec:opt_params}
In order to ascertain the optimal parameters, there are several approaches that may be adopted.
Firstly, there is the possibility of inspiration by adiabatic time-evolution as evidenced in the works of~\cite{montanez_barrera_towards_2024,zhou_quantum_2020}.
Secondly, there is the option of utilising classical optimisation routines~\cite{powell94, periyasamy24}, where parameter vectors \(\vec{\gamma}, \vec{\beta}\) are obtained via an iterative quantum-classical scheme.
In the standard formulation of \ac{qaoa}, the depth \(p\) is selected in advance.
In essence, larger values of \(p\) yield monotonically improving solution quality on perfect hardware.
However, for noisy systems, a trade-off exists between enhanced solution quality and an escalating amount of noise and imperfections caused by deeper circuits with growing \(p\) which has to be taken into account.
In the simplest case, the initial set of values for \(\vec{\gamma}, \vec{\beta}\) may be chosen randomly, although more informed choices are possible (see, \eg,~\cite{egger21}).

The \FOURIER strategy introduced by Zhou~\etal~\cite{zhou_quantum_2020} comprises two heuristic improvements to the basic scheme: (1) By executing multiple runs of the algorithm iteratively with growing values for \(p\), good initial estimates for \(\vec{\gamma}, \vec{\beta}\) are determined.
The optimal parameters obtained in run \(p\) are used to provide suitable initial values for the deeper circuit \(p+1\).
(2) To reduce the effective dimension of the parameter space, the set of \(2p\) parameters \((\vec{\gamma}, \vec{\beta}) \in \mathbb{R}^{2p}\) is replaced by a new set \((\vec{u}, \vec{v}) \in \mathbb{R}^{2q}\) with \(q\leq p\) so that each of the elements \(\gamma_{i}, \beta_{i}\) of the former set (with \(i\in[0,p-1]\)) can be expressed as a discrete sine/cosine transform of the set \(\vec{u}, \vec{v}\).
By choosing a specific value of \(q\), the dimension of the effective optimisation parameter space can be delimited at will, at the possible expense of solution quality, but also a reduced complexity of the classical optimisation sub-task of \ac{qaoa}.

\section{Methodology}
\label{sec:trackrec}
In this work, we address the \textbf{NP} problem of \acf{tr}, a key computing challenge in \ac{nhep}.
While several classical heuristic methods~\cite{atlas_collaboration_fast_2019,bocci_heterogeneous_2020} have been proposed to recognise tracks in the event data of the \ac{lhc}, recent publications also approach this problem from the perspective of quantum computing~\cite{tuysuz_particle_2020,zlokapa_charged_2021}, for instance through casting it in \ac{qubo} form~\cite{bapst_pattern_2019,schwagerl_particle_2023}.

The \ac{qubo} formulation that is employed in this work is introduced by Bapst~\etal~\cite{bapst_pattern_2019}, and also based on the corresponding implementation~\footnote{\href{https://github.com/derlin/hepqpr-qallse}{github/derlin/hepqpr-qallse}}.
While a comprehensive formulation is provided in Ref.~\cite{bapst_pattern_2019}, the central concept is to reconstruct complete tracks from smaller track segments consisting of three hit-points, referred to as \emph{triplets}.
The \ac{qubo} is then given by
\begin{equation}
  Q(\vec{a}, \vec{b}, \vec{T}) = \sum_{i=1}^{N} a_i T_i + \sum_{i=1}^{N} \sum_{j<i}^{N} b_{ij} T_i T_j,
\end{equation}
where $\vec{a} \in \mathcal{R}^N$ and $\vec{b} \in \mathcal{R}^{N \times N}$ are constant bias and coupling weights, respectively, and $\vec{T} \in \{0,1\}^N$ denotes variables for potential triplets.

With growing problem size, the \ac{qubo} formulation becomes too large to fit on current quantum systems, requiring one qubit per variable, and therefore also unfeasible to simulate classically.
This limitation is particularly evident in the case of full \ac{tr}, where the number of potential triplets increases rapidly with the complexity of the data.
To address this challenge, we adopt a strategy, similar to the approach in Ref.~\cite{schwagerl_particle_2023}, where we only focus on angle segments of a carefully selected data fraction~\cite{bapst_pattern_2019} of the hitpoints in the detector for the \ac{tr} process.
This approach also allows to obtain multiple problem instances from the data of one event.

In particular, for our numerical experiments, we filter $10\%$ of the TrackML data~\cite{Kiehn_2019} for one event, and divided it separately into $32$ and $64$ angle segments.
Additionally, we filtered $20\%$ of data for one event, and obtained 64 angle segments.
As we are simulating the quantum systems classically, we discard all \acp{qubo} encompassing more than $23$ variables, leading to $98$ problem instances in total.
While only a part of the tracks can be analysed this way, such reduced problems can act as a proof-of-concept on the potential of \ac{qa} and \ac{qaoa} approaches on \ac{tr}.

\section{Experimental Results}
\label{sec:exp}

\subsection{Spectral Gap Analysis}

\begin{figure}[ht]
  \includegraphics{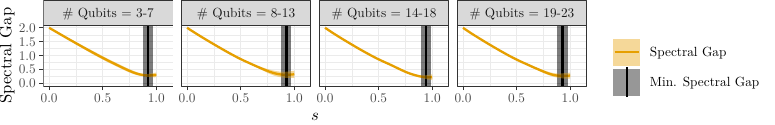}
  \caption{
    Average \ac{sg} and the corresponding position of the minimum \ac{sg} over interpolation parameter $s \coloneqq t/T$ for $98$ \ac{tr} problem instances with varying sizes. In each qubit range we computed \acp{sg} for $20\dots30$ problem instances. Shaded areas represent the standard deviation.
  }
  \label{fig:gap_stat}
\end{figure}

As described in \autoref{sec:qa}, the \ac{sg} gives an indication of the required runtime of an algorithm.
For the \ac{tr} instances considered in this work, we obtain the \acp{sg} of the interpolation from $\hat{H}_0$ to $\hat{H}_C$ by analytically computing the eigenvalues of \autoref{eq:QA} with 50 linearly increasing values of the interpolation factor $s \in [0,1]$ (\ie at $s=0$ and $s=1$, only $\hat{H}_0$ and $\hat{H}_C$ are represented, respectively).
As shown in \autoref{fig:gap_stat}, the average \ac{sg} does not exhibit significant variation across different qubit numbers, demonstrating a notable consistency.
This stability suggests that the system's energy landscape remains relatively predictable and stable as the problem size changes, indicating that annealing schedules relying on these \acp{sg} might maintain their performance characteristics regardless of the problem size.
This could be advantageous for tasks like \ac{tr}, where scalability is paramount.

\subsection{Annealing Schedules from \ac{qaoa}}

Given the similarity of the \acp{sg} throughout the considered problem instances, we randomly pick one $11$-qubit instance, for which we derive annealing schedules using \ac{qaoa} as a proof-of-concept.
The first two discrete energy levels and the \ac{sg} for this specific instance are shown in \autoref{fig:energy_selected}.

As described in \autoref{sec:opt_params}, we use the \FOURIER strategy~\cite{zhou_quantum_2020} to obtain near-optimal parameters.
To this end, we sequentially optimise the set of \FOURIER parameters $(\vec{u}, \vec{v}) \in \mathbb{R}^{2q}$ starting from \ac{qaoa} depth $p=1$ up to $p=50$ using the numerical \emph{COBYLA} optimiser~\cite{powell94}, and re-use the optimised parameters from each previous depth $p-1$ as initial values padded with zeros.
We employ different values for $q \leq q_\text{max}$, while $q \leq p$ to limit the number of frequencies in the course of the $(\vec{\beta}, \vec{\gamma}) \in \mathbb{R}^{2p}$ parameters.
Additionally, we directly optimise $\vec{\beta}, \vec{\gamma}$, starting from $p=1$ with a \ac{ri}, up to $p=50$, also re-using parameters from the previous depth $p-1$.

\begin{figure}[ht]
  \begin{minipage}[c]{0.5\textwidth}
    \includegraphics[width=\textwidth]{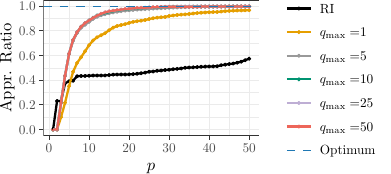}
  \end{minipage}\hfill
  \begin{minipage}[c]{0.45\textwidth}
    \caption{Energy approximation ratio obtained via \ac{qaoa} optimisation for increasing values of $p$.
      The approximation ratio curves for $q_\text{max} \in \{5, 10, 25, 50\}$ exhibit a significant overlap.}
    \label{fig:expval}
  \end{minipage}
\end{figure}

As demonstrated in \autoref{fig:expval}, the obtained approximation ratios (\ie the ratio between achieved energy and optimal energy) rapidly approach the optimum, when employing the \FOURIER initialisation.
Furthermore, a low-dimensional approximation of the parameter space is sufficient.
The use of \ac{ri} results in relatively lower approximation ratios, which indicates that the \FOURIER landscape is also easier to optimise classically.

\begin{figure}[ht]
  \begin{subfigure}{0.5\textwidth}
    \includegraphics[width=0.95\textwidth]{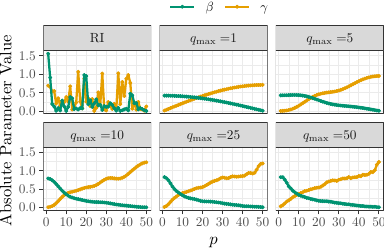}
  \end{subfigure}
  \begin{subfigure}{0.5\textwidth}
    \includegraphics[width=0.95\textwidth]{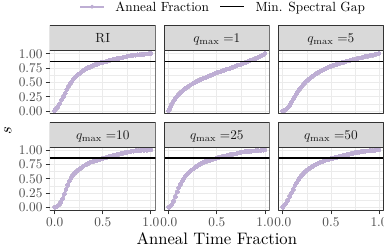}
  \end{subfigure}
  \caption{
    Left: Absolute parameter values $\vec{\beta}$ and $\vec{\gamma}$ obtained from the \ac{qaoa} optimisation at $p = 50$ for the \ac{tr} problem instance considered in this work.
    Right: Annealing schedules obtained from these parameters. The horizontal line denotes the position of the minimal \ac{sg}.
  }
  \label{fig:annealing-schedules}
  \vspace*{-1em}
\end{figure}

Since the parameter vectors \(\vec{\gamma}, \vec{\beta}\) can be derived from the effective parameters $\vec{u}, \vec{v}$, it is also possible to derive annealing schedules from these quantities, as described in \autoref{sec:anneal_schedules}.
The derived schedules are shown on the right side of \autoref{fig:annealing-schedules}, for the \ac{qaoa} depth $p=50$ and varying degrees of approximation (specified by \(q_\text{max}\)).
The left side of \autoref{fig:annealing-schedules} shows the \ac{qaoa} parameter vectors \(\vec{\gamma}, \vec{\beta}\), which were used for the schedule computation.
For \ac{ri}, the parameter values exhibit significant fluctuations compared to those obtained from the \FOURIER strategy.
However, the corresponding annealing schedule aligns with the intuition gained in Section~\ref{sec:qa} as annealing progresses rapidly until \(s\approx0.85\), as illustrated by the horizontal line, and subsequently slows down.

The \FOURIER strategy, which achieves near-optimal parameters even with $q_\text{max} = 5$ (cf. \autoref{fig:expval}), results in smooth parameter trajectories that produce similar schedules as \ac{ri} for $q_\text{max} \geq 5$.
Using too few effective \FOURIER parameters (\ie $q_\text{max} = 1$) leads to a suboptimal annealing schedule, which is closer to the default linear schedule.

\section{Discussion \& Conclusion}
\label{sec:concl}
The interplay between \ac{qaoa} parameters and annealing schedules presents a challenging \emph{chicken-and-egg} problem:
Achieving optimal \ac{qaoa} parameters can inform effective annealing schedules, while simultaneously, annealing motivates \ac{qaoa} in the first place.
This intricate relationship underscores the importance of balancing computational complexity with solution quality, particularly in the context of \ac{nisq} computing.

Our analysis has demonstrated that heuristics, such as the \FOURIER-based approach proposed in Ref.~\cite{zhou_quantum_2020}, offer valuable intuition and practical benefits for navigating this complex landscape.
By focusing on lower-frequency components in the parameter space, we have shown that it is possible to derive reasonable annealing schedules, reducing resource requirements while maintaining high solution quality.
While the implications of these findings and experimental validation of the obtained schedules on quantum annealers such as D-Wave systems extend beyond the specific context of the \ac{tr} problem, considered in this work, this study suggest promising avenues for generalisation across a wide range of \textbf{NP} optimisation problems within \ac{nhep} applications.
As co-designed quantum algorithms continue to evolve, the development of heuristic methods that bridge the gap between parameter optimisation and annealing schedules will play a pivotal role in unlocking quantum advantage.

The aim of this research is to contribute to a growing understanding of the interplay between \ac{qaoa} parameters and annealing schedules, as demonstrated on a specific problem in \ac{tr}.
As the field of quantum computing continues to advance, particularly from the \ac{nisq} era to fault-tolerance, such investigations will be instrumental in realising the full potential of quantum optimisation techniques to tackle complex, large-scale problems in diverse domains.

\bibliography{references}

\vspace{0em}\begin{small}
  \noindent\textbf{Acknowledgements}
  MF, MaSc and WM acknowledge support by the German Federal Ministry of Education and Research (BMBF), funding program 'quantum technologies—from basic research to market', grant numbers 13N16092 and 13N15647. WM acknowledges support by the High-Tech Agenda Bavaria.
  MD was supported by the U.S. Department of Energy Office of Science, Office of Nuclear Physics contract number DE-AC05-06OR23177, under which Jefferson Science Associates, LLC operates Jefferson Lab.
  MeSt, EK and AS acknowledge support by the state of Baden-Württemberg through bwHPC.
  PZ is funded by the \enquote{Atracción de Talento} Investigador program of the Comunidad de Madrid (Spain) No. 2022-T1/TIC-24024.
\end{small}

\end{document}